\documentclass[aps,pre,reprint,longbibliography,floatfix]{revtex4-2}
\usepackage[utf8]{inputenc}
\usepackage{amssymb,amsmath}
\usepackage{epsfig}
\usepackage{bm}
\usepackage{soul}
\usepackage{color}
\usepackage[mathlines]{lineno}%
\usepackage{epsfig}
\usepackage{comment}

\usepackage{graphicx}

\usepackage{siunitx}

\usepackage{dcolumn}

\newcommand\beq{\begin{equation}}
\newcommand\eeq{\end{equation}}
\newcommand\beqa{\begin{eqnarray}}
\newcommand\eeqa{\end{eqnarray}}

\newcommand{\rb}{\mathbf{r}}
\newcommand{\vb}{\mathbf{v}}
\newcommand{\qb}{\mathbf{q}}
\newcommand{\Fb}{\mathbf{F}}

\begin{document}

\title{  Particle-scale structure of granular suspensions
}

\author{Santos Bravo Yuste}
\email{santos@unex.es}
\homepage{https://fisteor.cms.unex.es/investigadores/santos-bravo-yuste}
\affiliation{Departamento de F\'{\i}sica and Instituto de Computaci\'on Cient\'{\i}fica Avanzada (ICCAEx), Universidad de Extremadura, E-06006 Badajoz, Spain}
\author{Antonio M. Puertas}
\email{apuertas@ual.es}
\affiliation{ Departamento de Qu\'{\i}mica y F\'{\i}sica, Universidad de Almer\'{\i}a, 04120 Almer\'{\i}a, Spain}

\begin{abstract}

Granular suspensions are intrinsically nonequilibrium systems in which dissipative grain-grain collisions coexist with solvent-induced forcing. We study the particle-scale structure of a granular suspension modeled by smooth inelastic hard spheres immersed in a thermal bath and compare Langevin-dynamics simulation results for the radial distribution function $g(r)$ and the static structure factor $S(q)$ with predictions of an equilibrium-inspired rational function approximation (RFA). The equilibrium hard-sphere RFA is supplied with nonequilibrium input for the contact value and a reduced isothermal-compressibility-like quantity, yielding analytical expressions for $g(r)$ in Laplace space  and  for $S(q)$. We find that the RFA gives a very good description of the short- and intermediate-range structure of the suspension over a broad range of densities, drag coefficients, and inelasticities. It reproduces $g(r)$ substantially better than the Percus-Yevick approximation in inelastic states, especially near contact, and gives a good account of $S(q)$ except at the smallest wave numbers. There, simulations show a drag-dependent enhancement over the RFA prediction, indicating additional long-wavelength nonequilibrium correlations beyond the present equilibrium-like description.  These results show that an equilibrium-based hard-sphere approach provides an accurate description of the particle-scale structure of the present Langevin model with inelastic hard spheres (except in the smallest-$q$ region), and suggest that similar equilibrium-inspired approaches may also be useful for related nonequilibrium hard-sphere suspension models, including multicomponent systems.

\end{abstract}

\date{\today}
\maketitle


\section{Introduction}
\label{sec1}

The statistical-mechanical determination of the structure of equilibrium fluids has been the subject of extensive theoretical, simulation, and experimental work for decades \cite{Hansen2013Book,Barker1976Review}. A simple model that nonetheless captures the basic phenomenology of these systems remarkably well is that in which molecules interact as elastic hard spheres. This model greatly simplifies the description of the fluid structure while still providing an accurate account of the equation of state, the fluid-solid transition, and the glass transition in systems such as uncharged colloidal suspensions. Nevertheless, an exact theoretical description of that structure is lacking, although accurate approximations have been developed.

A more recent extension of this model, which has also attracted enormous attention, is obtained when the particles are ``grains,'' i.e., particles that interact inelastically. Inelastic collisions are the defining ingredient of a granular gas  \cite{Brilliantov2004Book,Garzo2019Book}. Because collisions are inelastic, granular gases continuously lose kinetic energy and, in the absence of external energy input, eventually come to rest. In many situations, however, the grains are immersed in a surrounding medium (the carrier fluid or \textit{bath}), for instance a molecular gas, which continuously transfers energy to them and compensates for the collisional energy loss. In this way, a nonequilibrium stationary state (NESS) is reached. The grains together with the carrier fluid in which they are immersed constitute a granular suspension, a class of systems relevant in many natural phenomena and industrial applications.

The structure of a fluid at the particle scale is encoded in the probability of finding particles in given spatial arrangements relative to one another. Understanding that structure is important for both equilibrium and transport properties. Moreover, since suspensions often serve as starting materials for solids, their internal organization can strongly influence properties such as the strength and durability of the final product \cite{Russel1989book}. In simple fluids, this structural information is commonly characterized in terms of structural correlation functions (SCFs), in particular the radial distribution function $g(r)$ and the static structure factor $S(q)$. These quantities are equally natural in granular suspensions, since they characterize spatial correlations over a broad range of length scales and provide direct contact with experiments, simulations, and theoretical approaches from liquid-state physics \cite{Puglisi2012,Puertas2026}. The difficulty is that granular suspensions are intrinsically nonequilibrium systems: dissipative grain-grain collisions coexist with solvent-induced drag and stochastic forcing, so that the stationary state is not determined by conservative interactions alone. As a consequence, the standard equilibrium framework relating structure to interparticle interactions in ordinary fluids cannot be transferred straightforwardly to this context.

At the particle level, a simple and widely used effective description of dilute suspensions at low Reynolds number is provided by  Langevin  dynamics  in which the action of the surrounding fluid is represented by a viscous drag term together with a stochastic white-noise force \cite{Graham2018Book,TeGrotenhuis1994,Garzo2012}. When particle-particle collisions are important, the suspended phase is naturally described by a kinetic equation with a Boltzmann or Enskog collision operator; in granular suspensions, this operator incorporates collisional inelasticity \cite{Puglisi2015Book,Hayakawa2017,Ruben2019}. Recently, Gómez-González and Garzó showed that, in the Brownian limit, the Langevin-like forcing can be derived from a more microscopic collisional description of grains immersed in a molecular gas \cite{Ruben2022JFM,Ruben2025}.

This Langevin model does not account for all dynamical features of real suspensions, in particular long-range hydrodynamic interactions, which can qualitatively affect dynamical observables \cite{Padding2006}. However, previous work has shown that equilibrium structural quantities of the particles such as the radial distribution function may be reproduced very accurately even when the detailed dynamics is simplified, in contrast to time-dependent quantities such as velocity autocorrelation functions \cite{Padding2006}. Motivated by this separation between structural and dynamical sensitivity, in this paper we use molecular-dynamics simulations of the inelastic Langevin model to obtain the SCFs of granular suspensions in a nonequilibrium steady state, and we ask to what extent these can be described by an equilibrium-like liquid-state approximation. Our aim is not to develop a full theory of the nonequilibrium structure of granular suspensions, but to determine to what extent a simple equilibrium-like approximation can account for their particle-scale structure. To this end, we develop a rational function approximation (RFA) \cite{Yuste1991,Yuste1998}, equivalent to a generalized mean spherical approximation, for the SCFs of granular suspensions. The validity and scope of this description are then assessed through comparison with simulation results within the Langevin model. We will find that the agreement is generally very satisfactory and in many cases far better than that obtained from the Percus-Yevick (PY) approximation.

Although several theoretical approaches to nonequilibrium granular suspensions have been developed, an analytical description of their particle-scale structural correlation functions is still lacking.  Fluctuating-hydrodynamic theories have successfully described the long-wavelength behavior of density and velocity structure factors and the associated nonequilibrium correlation lengths \cite{Gradenigo2011,Puglisi2012}, but they do not provide an analytical description of the particle-scale density structure that is the main concern here. Likewise, in the one-dimensional study of Cecconi \textit{et al.} \cite{Cecconi2004}, the static structure factor of the inelastic system was compared with the corresponding elastic expression, rather than by means of a distinct theory for the inelastic case. More generally, simulation and experimental information on SCFs in granular suspensions remains comparatively limited \cite{Cecconi2004,Puglisi2012,Puertas2026}.
Besides particle-resolved imaging, acoustic propagation may provide an indirect measurement of SCFs, as recently shown by Tourin et al.  \cite{Tourin2026}.

The remainder of the paper is organized as follows. Section~\ref{sec:Langevin} introduces the Langevin model of the granular suspension considered here and describes the molecular-dynamics simulation procedure. In Sec.~\ref{sec:gsigmaS0}, we provide approximate expressions for the contact value of the radial distribution function and the isothermal compressibility of the suspension. These quantities are required as inputs for the RFA model, proposed in Sec.~\ref{sec:RFA}, to account for the structure of the granular suspension. Section~\ref{sec:results} presents a comparison between the radial distribution function and the static structure factor obtained from the RFA model and the PY approximation, on the one hand, and the simulation results, on the other, for different values of the density, drag coefficient, and collision inelasticity. Finally, the paper ends with a section summarizing the main findings, presenting our conclusions, and pointing to a possible extension of the present work to multicomponent granular suspensions.

  \section{Langevin model of a granular suspension}
\label{sec:Langevin} 

In the Langevin description of a granular suspension, the velocity $\vb_i$ of particle $i$ satisfies \cite{Puglisi2005,Gradenigo2011,Garzo2012,Hayakawa2017,Garzo2019Book}
\begin{equation}
	m \dot \vb_i=-\gamma \vb_i +\Fb_i^\text{st}+\Fb_i^\text{c},
	\label{ecuLangevin}
\end{equation}
where $m$ is the particle mass, $\vb_i$ its velocity, and $\gamma\ge 0$ the drag or friction coefficient associated with the background fluid (bath). The second term, $\Fb_i^\text{st}$, is a stochastic force, taken as Gaussian white noise with the properties $\langle \Fb_i^\text{st}\rangle=0$ and $\langle \Fb_i^\text{st}(t)\Fb_j^\text{st}(t')\rangle=2\gamma k_\text{B} T_\text{b} \mathbf{I} \delta_{ij} \delta(t-t')$, where $\mathbf{I}$ is the unit tensor, $T_\text{b}$ is the bath temperature, and $k_\text{B}$ is the Boltzmann constant.
The term $\Fb_i^\text{c}$  accounts for binary 
collisions between particles, which are modeled as smooth hard spheres of diameter $\sigma$. {Note that the stochastic force acts as an energy source that keeps the system in motion and avoids the clustering instability typical of granular gases \cite{Puertas2026}.} 

If particles $i$ and $j$ collide, their precollisional and postcollisional velocities satisfy $\alpha\hat{\mathbf{\sigma}}\cdot (\vb'_i-\vb'_j)=\hat{\mathbf{\sigma}}\cdot (\vb_i-\vb_j)$, where primes denote precollisional velocities, unprimed symbols denote postcollisional velocities, and $\hat{\mathbf{\sigma}}$ is the unit vector joining the centers of the two colliding particles. The coefficient of (normal) restitution $\alpha$, with $0\le \alpha\le 1$, quantifies the degree of collisional inelasticity: the collision is elastic when $\alpha=1$ and inelastic otherwise.

When $\gamma=0$, Eq.~\eqref{ecuLangevin} reduces to $m \dot \vb_i=\Fb_i^\text{c}$, corresponding to the dry limit, i.e., to a granular gas. We will also refer to this case as the dry case. Here, the kinetic energy of the particles (their ``temperature'') decreases with each inelastic collision. If the system remains spatially homogeneous, it evolves in the homogeneous cooling state (HCS), for which the granular temperature decays according to Haff's law \cite{Haff1983,Brilliantov2004Book,Garzo2019Book}. By contrast, $\gamma\neq 0$ defines the wet case, namely the granular suspension. The combination of the friction and random forces acts effectively as a thermostat at temperature $T_\text{b}$, providing energy to maintain the system in a stationary state.
For $\alpha<1$, the inelasticity of the collisions provides an additional dissipative mechanism besides viscous drag. Hence, in the stationary state, the granular temperature satisfies $T<T_\text{b}$: the bath injects energy into the grains to compensate for the losses due to drag and inelastic collisions.

\subsection{Simulation details}

In the simulations, a system of $N=8000$ identical hard spherical particles is considered in a cubic box with periodic boundary conditions. The particle mass $m$, diameter $\sigma$, and bath thermal energy $k_\text{B}T_\text{b}$ are taken as units of mass, length, and energy, respectively. The equation of motion for each particle, Eq.~\eqref{ecuLangevin}, is solved using a Heun algorithm with a time step of $2.5\cdot 10^{-3}\,\sigma \sqrt{m/k_\text{B}T_\text{b}}$. Collisions between particles are identified by overlaps and handled using the conventional rules for smooth spherical particles (see, e.g., \cite{Brilliantov2004Book}). {The numerical error introduced by this algorithm is of the order of $\delta t$, which is much smaller than the random displacement due to the Brownian force for our time step. Indeed, simulations with smaller time steps showed no noticeable effect on the structure.}

Three different values of the particle friction coefficient with the background fluid were used in this work, $\gamma=\gamma^*(m k_\text{B}T_\text{b})^{1/2}/\sigma$  with $\gamma^*=0.2,1,5$.
These values span weak, intermediate, and strong coupling to the bath, and, as we will show in Sec.~\ref{sec:results}, are sufficient to reveal the influence of $\gamma$ on the structure of the granular suspension. Note that $\gamma^*\approx 1$ marks the frontier between diffusive to ballistic regimes in the granular suspension. The density, reported as volume fraction $\phi=\rho \pi \sigma^3/6$, with $\rho$ the number density, was varied between $\phi=0.10$ and $0.40$. 

Finally, the restitution coefficient entering the collision rules was varied between $\alpha=1.0$ (elastic spheres) and $\alpha=0.3$. For the calculation of the SCFs, 500 independent configurations were extracted from a single simulation; successive configurations are separated by a mean squared displacement of 
$1.25 \sigma^2/\gamma^*$. 
Every state was prepared from scratch and aged for $1000$ time units before the particle  trajectories were recorded. It was found that this time was long enough to reach stationarity in all dynamic and static observables, in particular the pair distribution function and the structure factor.

For the structure factor, only wavevectors compatible with the periodic boundary conditions are considered, 
$\qb = \dfrac{2\pi}{L} (n_x, n_y, n_z)$, 
where $L$ is the simulation box size and $n_x$, $n_y$, and $n_z$ are integers. {In order to estimate the uncertainty in $S(q)$, the whole set of $500$ configurations has been divided into $M=5$ subsets.  In each subset, the average structure factor $S_i(q)$ is calculated, and the final $S(q)$ is taken as the mean of $S_i(q)$, and the error as the standard error of the mean, namely, for every wavevector $q$,}

\[ \Delta S(q) = \sqrt{\frac{\sum_{i=1}^M \left[S_i(q) - S(q) \right]^2}{M (M-1)}}, \]

\noindent {For the pair distribution function, the error bars calculated in this way are smaller than the symbol size in the figures of Sec.~\ref{sec:results} and are not shown.}

{A finite-size analysis was also performed by comparing simulation results from two systems with $N=1000$ and $N=8000$ particles. While both systems show similar results, the larger system allows reaching smaller wavevectors and reduces the statistical uncertainty.}

\section{Radial distribution function at contact and isothermal compressibility for a granular suspension}
\label{sec:gsigmaS0}

To apply the RFA beyond the PY approximation, one needs as input the radial distribution function at contact of the grains and the isothermal compressibility of the granular suspension. Approximate expressions for these quantities are given below.

\subsection{Radial distribution function at contact}
\label{rdfContac}

Let $g_0(\mathbf{r}_1,\mathbf{r}_2)$ be the precollisional pair distribution function for two granular particles at contact. Its spatial average under the condition $|\mathbf{r}_1-\mathbf{r}_2|=\sigma$ is denoted, following Lutsko \cite{Lutsko2001b}, by $g_0(\sigma)$. This quantity is also often denoted by $\chi$ or $\chi_0$ \cite{Garzo2007,Lutsko2001b}.

For the HCS of a granular gas (or dry granular suspension, $\gamma=0$), Lutsko \cite{Lutsko2001b} derived, within the Enskog approximation, the relation
\begin{equation}
	\label{gg0}
	g(\sigma)=\frac{1+\alpha}{2\alpha}\,g_0(\sigma).
\end{equation}
Equation \eqref{gg0} is the monocomponent limit of the mixture relation given by equation (3.7) of \cite{Garzo2007}.

The derivation of Eq.~\eqref{gg0} given in the appendix of \cite{Lutsko2001b} or in Appendix A of \cite{Garzo2007} applies directly to the present granular suspension model because the arguments used in those references—the hard-sphere collisional boundary condition at contact and the Enskog factorization of the precollisional two-body distribution—are exactly the same when $\gamma\neq 0$. In the present model, the drag and stochastic terms modify the between-collision dynamics but not the instantaneous binary collision rule. Therefore, the same approximate contact relation, Eq.~\eqref{gg0}, applies to granular suspensions as well.

To close Eq.~\eqref{gg0}, one must provide the value of $g_0(\sigma)$. The usual approximation for the HCS of a granular gas is to replace $g_0(\sigma)$ by the value of the pair distribution function of an elastic gas at equilibrium at the same density: $g_0(\sigma)=g_\text{eq}(\sigma)$. Accordingly, for the granular suspension within the Langevin model, Eq.~\eqref{gg0} becomes
\begin{equation}
	\label{gsgeq}
	g(\sigma;\phi,\alpha,\gamma)=\frac{1+\alpha}{2\alpha}\,g_\text{eq}(\sigma;\phi).
\end{equation}
Recall that $\phi$ is the volume fraction occupied by the grains. This equation has already been used in Ref.~\cite{Lutsko2001b} for the HCS of a monocomponent granular gas. The corresponding extension to mixtures is given in Ref.~\cite{Garzo2007}.

Equation \eqref{gsgeq} arises from an Enskog-type closure that replaces the true precollisional contact correlations by equilibrium-like ones and neglects velocity correlations generated by inelasticity and by the Langevin forcing. In other words, $g(\sigma)$ is expected to depend on $\gamma$, even though this dependence is absent from the approximate closure \eqref{gsgeq}. In Sec.~\ref{sec:results} we will show that the simulation results indeed display a dependence on $\gamma$.

\subsection{Isothermal compressibility}

As in Refs.~\cite{Lutsko2001b,Ruben2019}, we adopt for the granular suspension the same compressibility factor as in the dry case:
\begin{equation}
	\label{Zgranu}
	Z=\frac{p}{\rho k_\text{B} T}
	=1+2(1+\alpha)\phi g_0(\sigma),
\end{equation}
where $\rho$ is the number density defined previously, $p$ is the granular pressure, and $T$ is the granular temperature. 
The derivation of this equation is the same as for the HCS of a dry granular system \cite{Garzo1999,Lutsko2001b}, since in our model the drag and stochastic forces do not modify the collisional contribution to the virial pressure. The expression of $Z$ for a multicomponent granular suspension is given in Ref.~\cite{Garzo2007}.

In Sec.~\ref{sec:RFA} we will see that the RFA requires as input the quantity playing the role of the reduced isothermal compressibility  (or isothermal susceptibility) $\hat{\kappa}=\rho k_\text{B} T \kappa_T$, where 
 $\kappa_T=-\frac{1}{V}\left[\partial V/\partial p\right]_{N,T}$ 
is the isothermal compressibility. (Here we use the symbol $\kappa$ for the isothermal compressibility instead of the usual $\chi$ in order to avoid confusion with the contact value of the pair correlation function, which is often denoted by $\chi$ in kinetic theory.) From the definition of $Z$, one obtains
\begin{equation}
	\label{chiTermoA}
	\hat{\kappa}^{-1}= \left[\frac{\partial (\phi Z)}{\partial \phi}\right]_{N,T}.
\end{equation}

If, as in Sec.~\ref{rdfContac}, we approximate $g_0(\sigma)$ by its equilibrium counterpart $g_{\mathrm{eq}}(\sigma)$, then
\begin{equation}
	\label{hatkappa}
	\hat{\kappa}^{-1}-1= \frac{1+\alpha}{2}\left[\hat{\kappa}^{-1}_\text{eq}-1\right],
\end{equation}
where $\hat{\kappa}_\text{eq}$ is the reduced isothermal compressibility of an elastic gas at the same density. This equation was already proposed in Ref.~\cite{Lutsko2001b}, although for the HCS of a granular gas.

\section{Structural correlation functions by means of the RFA}
\label{sec:RFA} 

Here we give a brief description of the RFA method for single-component hard-sphere fluids at equilibrium. The RFA can be regarded as an alternative formulation of the generalized mean spherical approximation \cite{Yuste1991}. Further details on the RFA method can be found in \cite{Santos2020,Santos2016book}.

The Laplace transform of $r g(r)$ is
\begin{equation}
G(s)=\int_0^\infty dr\, r\, g(r) e^{-s r},
\end{equation}
and the structure factor $S(q)$ is defined by
\begin{equation}
	S(q)=1+\rho\int d\rb\,e^{i\mathbf{q}\cdot \rb} \, [g(r)-1].
\end{equation}
For an isotropic fluid, this can be rewritten as
\begin{equation}
	\label{Sq}
	S(q)=1-2\pi \rho \, \left. \frac{G(s)-G(-s)}{s}\right|_{s=iq}.
\end{equation}
The isothermal susceptibility is given by \cite{Evans1978}
\begin{equation}
	\label{kappaS0}
	\hat{\kappa}=1+\rho\int d\rb  \, [g(r)-1]=S(0).
\end{equation}
{ Equation~\eqref{kappaS0} relates $\hat{\kappa}$ (the reduced mechanical compressibility obtained from the approximate equation of state) to $S(0)$, the $q\to 0$ limit of the structure factor, which measures long-wavelength density fluctuations. In equilibrium, both quantities coincide, i.e.,  Eq.~\eqref{kappaS0}  becomes an exact identity, but this is not guaranteed in a NESS. Thus, testing to what extent $\hat{\kappa}$ is close to $S(0)$ in our NESS also probes how well the suspension behaves like an effective equilibrium system at long wavelengths. Note, however, that part of this difference may also arise from the fact that $\hat{\kappa}$, as given by Eq.~\eqref{hatkappa}, is itself approximate.}

In the RFA for hard spheres the function $G(s)$ takes the form (here and in what follows we take $\sigma=1$):
\begin{equation}
	\label{Gs}
	G(s)=\frac{F(s)e^{-s}}{1+12\phi F(s) e^{-s}}
\end{equation}
where $F(s)$ is a rational function:
\begin{equation}
	F(s)=-\frac{1}{12\phi}\, \frac{1+L_1 s+L_2 s^2}{S_0+S_1 s+S_2 s^2+S_3 s^3+S_4 s^4}.
\end{equation}

If one sets $L_2=S_4=0$, the remaining coefficients $\{L_1,S_1,S_2,S_3\}$ are uniquely determined by requiring that both the contact value $g(\sigma)$ and the  isothermal susceptibility $\hat{\kappa}$ remain finite. The resulting coefficients are precisely those leading to the PY approximation for hard spheres \cite{Yuste1991}. It is well known that the PY approximation yields expressions for $g(\sigma)$ and $\kappa_T$ that suffer from thermodynamic inconsistency: the compressibility factor $Z$ obtained via the virial route does not coincide with that obtained via the compressibility route \cite{Santos2016book}. This inconsistency can be corrected by choosing appropriate values of $L_2$ and $S_4$. In other words, one can select the values of $L_i$ and $S_i$ so as not only to satisfy the basic finiteness conditions for $g(\sigma)$ and $\hat{\kappa}$, but also to reproduce prescribed values of $g(\sigma)$ and $\hat{\kappa}$. The final expressions are 
\begin{align}
	S_0&=1,            \\
	S_1&= -1+L_1,           \\
	S_2&= \frac{1}{2}-L_1+L_2,         \\
	S_3&= -\frac{1+2\phi}{12\phi}+\frac{L_1}{2}-L_2,\\
	S_4&=  \frac{1+\phi/2}{12\phi}-\frac{1+2\phi}{12\phi}L_1+\frac{L_2}{2}, 
\end{align}
where
\begin{equation}
	\label{L2}
	L_2=\frac{-1-\phi/2+(1+2\phi)L_1}{1+6\phi g(\sigma)} g(\sigma) 
\end{equation}
and
\begin{equation}
	\label{L1}
	L_1=\frac{1}{2}
	+\left\{\frac{1}{12\phi} 
	\frac{(1-\phi)^2-\left[6\phi g(\sigma)+1\right]\hat{\kappa}}
	{2+\phi-2(1-\phi)^2g(\sigma)} 
	\right\}^{1/2}.
\end{equation}
We have written the above relations in the form used by Lutsko in Ref.~\cite{Lutsko2001b} in order to make clear the similarity between our approach and the one proposed in that work.

In this work, we employ the above RFA expressions, originally derived for equilibrium hard-sphere fluids, as an approximate representation of the structural functions $g(r)$ and $S(q)$ of granular suspensions. Specifically, the quantities inserted into Eqs.~\eqref{L2} and \eqref{L1} are the granular-suspension expressions for $g(\sigma)$ and $\hat{\kappa}$, given by Eqs.~\eqref{gsgeq} and \eqref{hatkappa}, respectively.
In these equations we will use the Carnahan–Starling values for the  equilibrium quantities:
\begin{align}
\label{geqCS}	g_\text{eq}(\sigma)&= \frac{1-\phi/2}{(1-\phi)^3},  \\
	\hat{\kappa}_\text{eq}^{-1}&=\frac{1+4\phi+4\phi^2-4\phi^3+\phi^4}{(1-\phi)^4}.
	\label{keqInv}	             	
\end{align}

Therefore, Eqs.~\eqref{Gs}--\eqref{keqInv}, combined with
Eq.~\eqref{Sq}, provide a fully analytical representation of the
radial distribution function of granular suspensions in Laplace
space, $G(s)$, together with the corresponding structure factor,
$S(q)$. The radial distribution function in real space, $g(r)$, can then be conveniently obtained by numerical inversion of $G(s)$ \cite{NumLaplaInv} or, analytically, for the first few shells, $\sigma \le r <n\sigma$ with $n$ a small integer~\cite{Yuste1991,Santos2020}.

The RFA method proposed here for determining the SCFs of a granular suspension can be viewed as an extension to the wet case ($\gamma \neq 0$) of the approach proposed by Lutsko in Ref.~\cite{Lutsko2001b} for the HCS in the dry case.

\section{Results: Comparison of theory and simulations}
\label{sec:results}

In this section, we compare molecular-dynamics simulation results with the predictions of the RFA and PY theories for $g(r)$ in Sec.~\ref{sec:gr} and for $S(q)$ in Sec.~\ref{sec:Sq}.  

As mentioned above, we have carried out molecular-dynamics simulations for $\alpha=0.3,0.4,\ldots,1$, $\phi=0.1,0.2,0.3,0.4$, and $\gamma^*=0.2,1,5$. In Ref.~\cite{datosEnZenodo} we provide all simulation data together with the corresponding RFA and PY predictions. Here, for brevity, we only present the cases with $\alpha=0.4,0.6,0.8,1$, $\gamma^*=0.2,5$, and $\phi=0.1,0.4$. These cases illustrate the effects of inelasticity, friction, and density on the SCFs, and allow us to assess the quality of the theoretical approximations.

\subsection{Pair distribution function}
\label{sec:gr}

\begin{figure}
	\begin{center}		\includegraphics[width=.98\columnwidth]{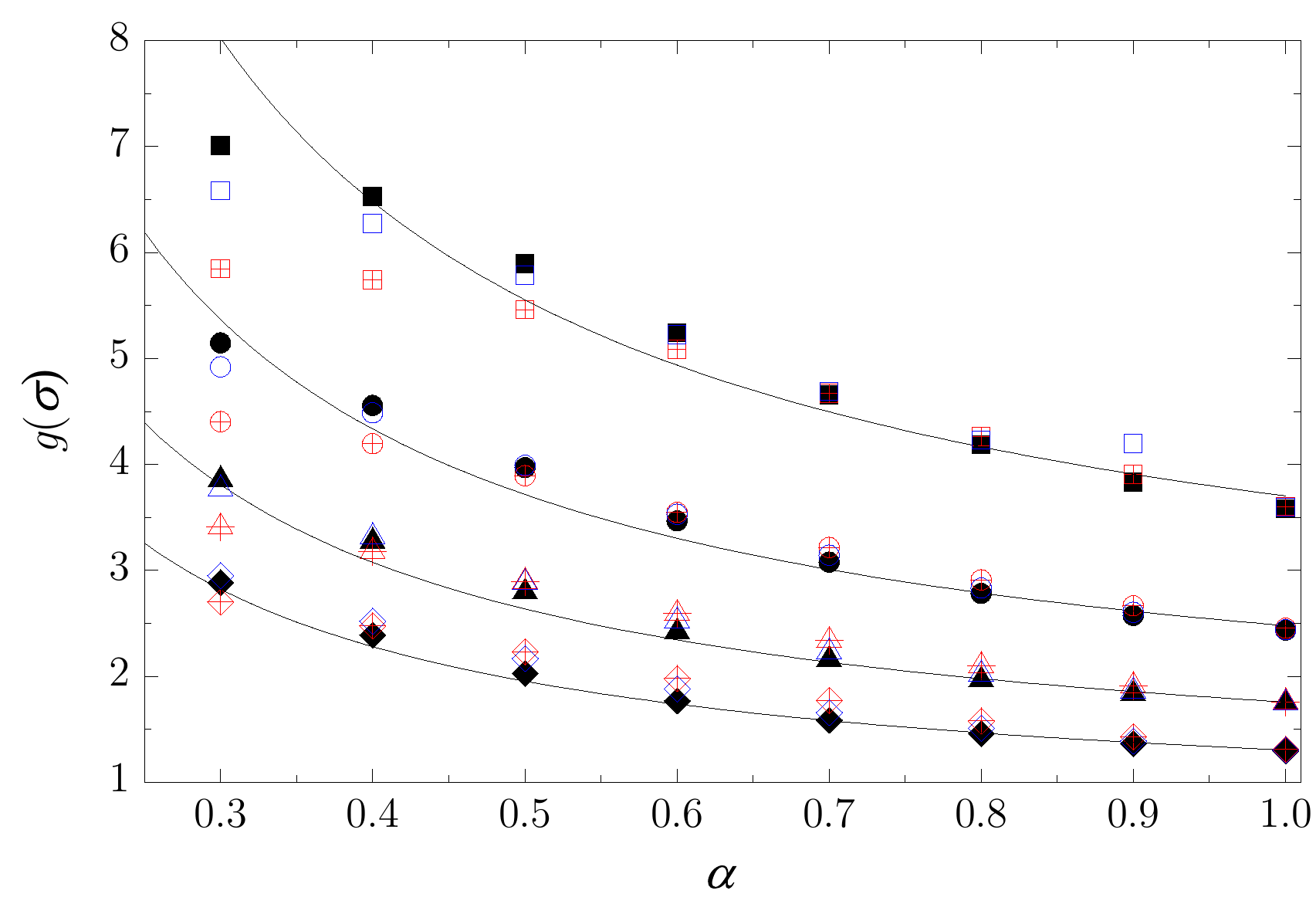}
	\end{center}
	\caption{ 
      Simulation values of $g(\sigma)$ versus the inelasticity coefficient $\alpha$ for $\phi = 0.1$ (diamonds), 0.2 (triangles), 0.3 (circles), and 0.4 (squares). The solid, open, and crossed symbols correspond to $\gamma^* = 0.2$, 1, and 5, respectively. The solid lines correspond to Eq.~\eqref{gsgeq} for $\phi = 0.1, 0.2, 0.3$, and 0.4, from bottom to top.}
	\label{fig:gSigmaBig}	
\end{figure}

As mentioned above, the contact value $g(\sigma)$ is a crucial input of the RFA theory, and therefore its accuracy strongly affects the quality of the RFA description. Figure~\ref{fig:gSigmaBig} shows the simulation results for $g(\sigma)$ together with the theoretical estimate obtained from Eqs.~\eqref{gsgeq} and \eqref{geqCS} for several volume fractions and reduced drag coefficients. The agreement is generally quite good, although noticeable deviations appear at the highest density and smallest values of $\alpha$. As expected, increasing inelasticity enhances $g(\sigma)$, reflecting a larger probability of finding particles in near-contact configurations. The simulations also display a dependence on $\gamma^*$, which is not contained in the theoretical estimate used as input in the RFA. 
A possible explanation is that the solvent affects not only the single-particle velocity distribution but also precollisional and near-contact velocity and positional correlations, an effect not accounted for in the approximation leading to Eq.~\eqref{gsgeq}. Because an inelastic collision reduces the outgoing normal component of the relative velocity by a factor $\alpha$, it tends to make the velocities of the colliding particles more nearly parallel than in the elastic case \cite{RubenRW2023,Yuste2026}. 
This collision-induced persistence can generate correlations between the velocities of nearby particles \cite{Brilliantov2004Book}. Drag and stochastic forcing relax and randomize particle velocities between successive collisions. As $\gamma$ increases, the bath-induced relaxation may weaken those correlations. This could explain the observed decrease of $g(\sigma)$   with increasing $\gamma$, which shifts it in the direction of its elastic value.

\begin{figure}
	\begin{center}		\includegraphics[width=.98\columnwidth]{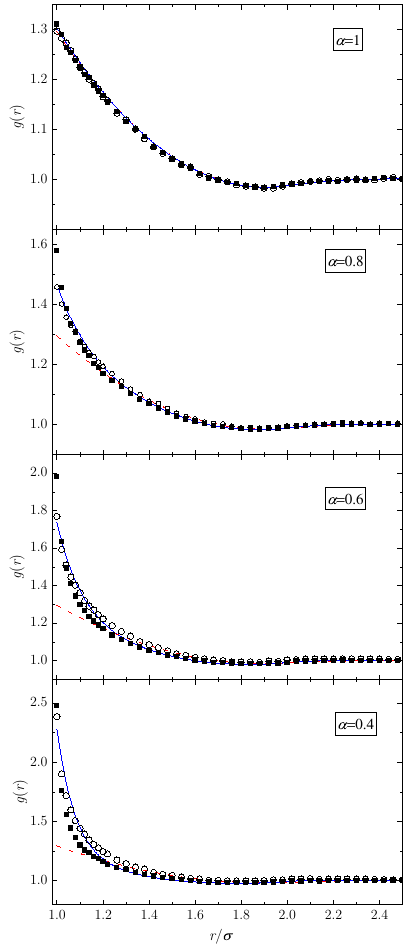}
	\end{center}
\caption{Radial distribution function $g(r)$ as a function of the scaled distance $r/\sigma$ for $\phi=0.1$. Symbols denote simulation results: circles for $\gamma^*=0.2$ and squares for $\gamma^*=5$. The PY and RFA predictions are shown by the red dashed and blue solid lines, respectively. Panels (a)--(d), from top to bottom, correspond to $\alpha=1$, $0.8$, $0.6$, and $0.4$. }
	\label{fig:grgamma02-5phi01}	
\end{figure}

\begin{figure}
	\begin{center}		\includegraphics[width=.98\columnwidth]{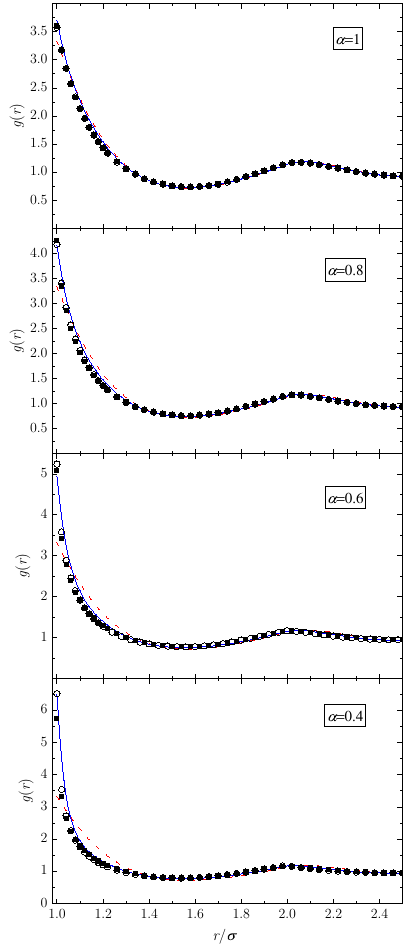}
	\end{center}
	\caption{Same as in Fig.~\ref{fig:grgamma02-5phi01}, but for $\phi=0.4$. 	}
	\label{fig:grgamma02-5phi04}	
\end{figure}

Figures~\ref{fig:grgamma02-5phi01} and \ref{fig:grgamma02-5phi04} show the results for the full $g(r)$ for $\phi=0.1$ and $\phi=0.4$, respectively, with two friction coefficients, $\gamma^*=0.2,5$, and four values of the coefficient of restitution, $\alpha=1, 0.8, 0.6,0.4$. 
As mentioned previously, since the theoretical predictions of PY and RFA do not depend on $\gamma$, the curves corresponding to these approximations are the same for $\gamma^*=0.2$ and $\gamma^*=5$.

As anticipated by the comparison of $g(\sigma)$, the main conclusion that can be drawn from Figs.~\ref{fig:grgamma02-5phi01} and \ref{fig:grgamma02-5phi04} (and is also supported by the other simulated cases; see Ref.~\cite{datosEnZenodo}) is that the quality of the theoretical description of $g(r)$ depends mainly on inelasticity and density, while the influence of $\gamma^*$, although weaker overall, becomes more visible as the density decreases.
Also we see that in the elastic case, $\alpha=1$, both PY and RFA reproduce the simulation data extremely well at both densities, although for $\phi=0.4$ the RFA is slightly more accurate near contact. Moreover, in this elastic regime the simulation results are practically independent of $\gamma$, so that drag has essentially no visible effect on $g(r)$.

Once inelasticity is introduced, however, the situation at low density, $\phi=0.1$ (see Fig.~\ref{fig:grgamma02-5phi01}), changes significantly. In this case, the simulation results do depend on $\gamma$, and this dependence becomes stronger as $\alpha$ decreases. The effect is concentrated mainly for short distances, $r\approx \sigma$, and, in fact, the strongest deviations are observed at the contact value $g(\sigma)$, with a clear dependence on the drag coefficient, as discussed in Figure \ref{fig:gSigmaBig}. The PY approximation deteriorates noticeably as inelasticity increases, especially near contact, whereas the RFA also worsens, although overall it still provides a reasonable description of the simulation data. Away from contact, the discrepancies between theory and simulation are much smaller.

At the higher density, $\phi=0.4$, (see Fig.~\ref{fig:grgamma02-5phi04}) the elastic case remains equally well described by both theories, and the simulation data are again essentially insensitive to $\gamma$. The important difference with respect to the case $\phi=0.1$ appears in the inelastic states: in clear contrast with the low-density situation, the simulation results now depend only very weakly on the drag coefficient. In practice, the two values of $\gamma^* $ produce almost indistinguishable curves, and only for the most inelastic cases does one observe a small difference, again localized near contact. 
As before for the low density case, the PY approximation worsens as $\alpha$ decreases, particularly in the vicinity of $r=\sigma$, while the RFA also degrades somewhat but still yields a reasonable overall description. In fact, for strong inelasticity the RFA performs better at high density than at low density.

Taken together, these results show that the influence of drag and stochastic forcing on the structure of the granular suspension is density dependent. 
A possible interpretation is that, at low density, the longer time between successive collisions gives the bath more time to affect the particle dynamics. 
Their influence may become more visible as inelasticity increases, because more inelastic collisions tend to produce larger departures of the particle velocities from those associated with the bath temperature. 
By contrast, at higher density the more frequent collisions would reduce the time over which the bath can affect the precollisional dynamics, so that the structure becomes controlled mainly by density and inelasticity, while the influence of $\gamma^*$ remains comparatively weak. In any case, the comparison clearly shows that the main limitation of both PY and RFA appears in the near-contact region as inelasticity increases, although the RFA remains systematically more accurate than PY, especially for $\alpha<1$.

\subsection{Structure factor}
\label{sec:Sq}

While $g(r)$ provides the most direct characterization of local packing and short-range order in real space, $S(q)$ shows how density fluctuations are distributed over different length scales in reciprocal (Fourier) space. In particular, $S(q)$ distinguishes particle-scale structure, encoded at large $q$, from collective long-wavelength fluctuations at small $q$, and is therefore especially useful for identifying nonequilibrium effects that may  be only weakly visible in $g(r)$.

\begin{figure}
	\begin{center}
		\includegraphics[width=.98\columnwidth]{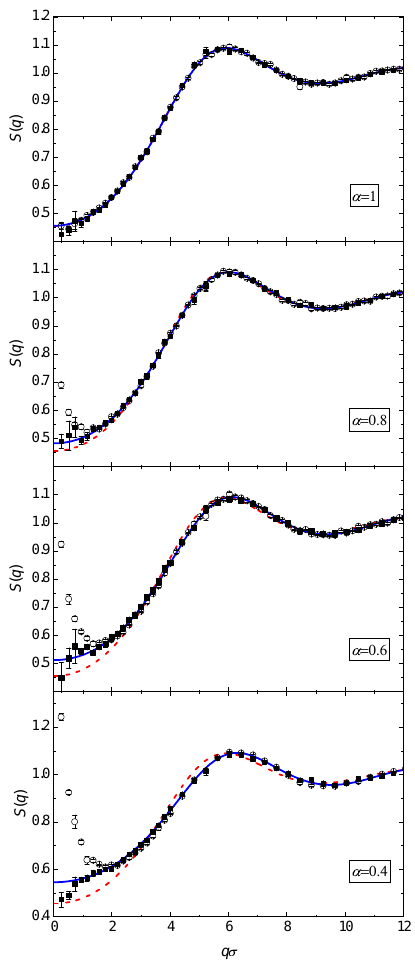}
	\end{center}
	\caption{Static structure factor $S(q)$ versus $q\sigma$ for $\phi=0.1$. Symbols denote simulation results: circles for $\gamma^*=0.2$ and squares for $\gamma^*=5$. The PY and RFA predictions are represented  by the red dashed and blue solid lines, respectively. Panels (a)--(d), from top to bottom, correspond to $\alpha=1$, $0.8$, $0.6$, and $0.4$.	}
	\label{fig:Sqgamma02-5phi01}	
\end{figure}

\begin{figure}
	\begin{center}    \includegraphics[width=.98\columnwidth]{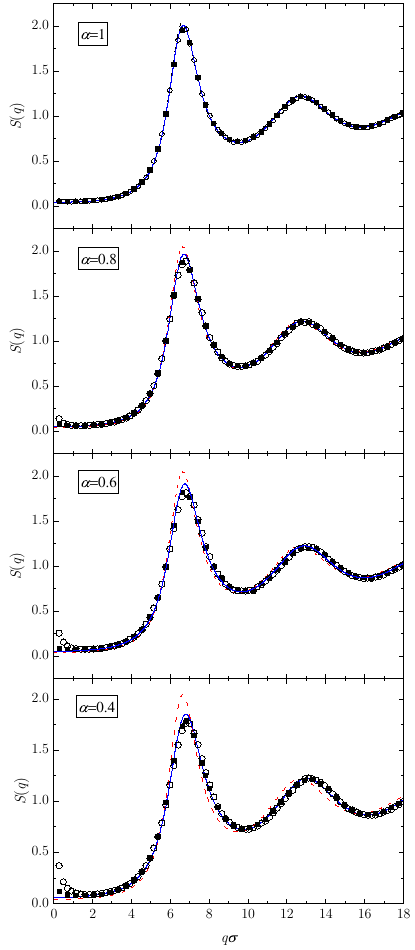}
	\end{center}
	\caption{Same as in Fig.~\ref{fig:Sqgamma02-5phi01}, but for $\phi=0.4$. {Error bars are not shown because they are smaller than the symbols.}	}
	\label{fig:Sqgamma02-5phi04}	
\end{figure}

Figures~\ref{fig:Sqgamma02-5phi01} and \ref{fig:Sqgamma02-5phi04} show the results for $S(q)$ for $\phi=0.1$ and $\phi=0.4$, respectively, with two friction coefficients, $\gamma^*=0.2,5$, and four values of the coefficient of restitution, $\alpha=1,0.8,0.6, 0.4$. 
Note that, since the theoretical predictions of PY and RFA do not depend on $\gamma$, the curves corresponding to these approximations are the same for $\gamma^*=0.2$ and $\gamma^*=5$. On the other hand, the simulations at low density and in the low-$q$ region show increased statistical errors because there are too few wavevectors compatible with the periodic boundary conditions,  reducing the averaging over equivalent wavevectors.

The results shown in the figures, as well as those corresponding to the additional cases provided in Ref.~\cite{datosEnZenodo}, reveal the same general trends as those observed for the radial distribution function.
In the elastic case, $\alpha=1$, both the PY approximation and the RFA provide an excellent description of the simulation data at both densities, while the simulations themselves are essentially independent of $\gamma$. This confirms that, in the elastic regime, the drag coefficient does not produce any appreciable effect on the structure factor. 

At low density, $\phi=0.1$, see Fig.~\ref{fig:Sqgamma02-5phi01}, the effect of inelasticity is much more pronounced. The simulation results become clearly dependent on $\gamma$, and the difference between the curves for different values of the drag coefficient increases as the inelasticity grows. This dependence is especially marked in the small-$q$ region, corresponding to long-wavelength density fluctuations. In particular, $S(0)$ exhibits a clear dependence on $\gamma$, which becomes stronger as $\alpha$ decreases. In this regime, the PY approximation deteriorates significantly, especially as $q \to 0$, whereas the RFA also loses accuracy but still yields, in general, a reasonable description of the simulation data except in the long-wavelength regime.

\begin{figure}
	\begin{center}
		\includegraphics[width=.98\columnwidth]{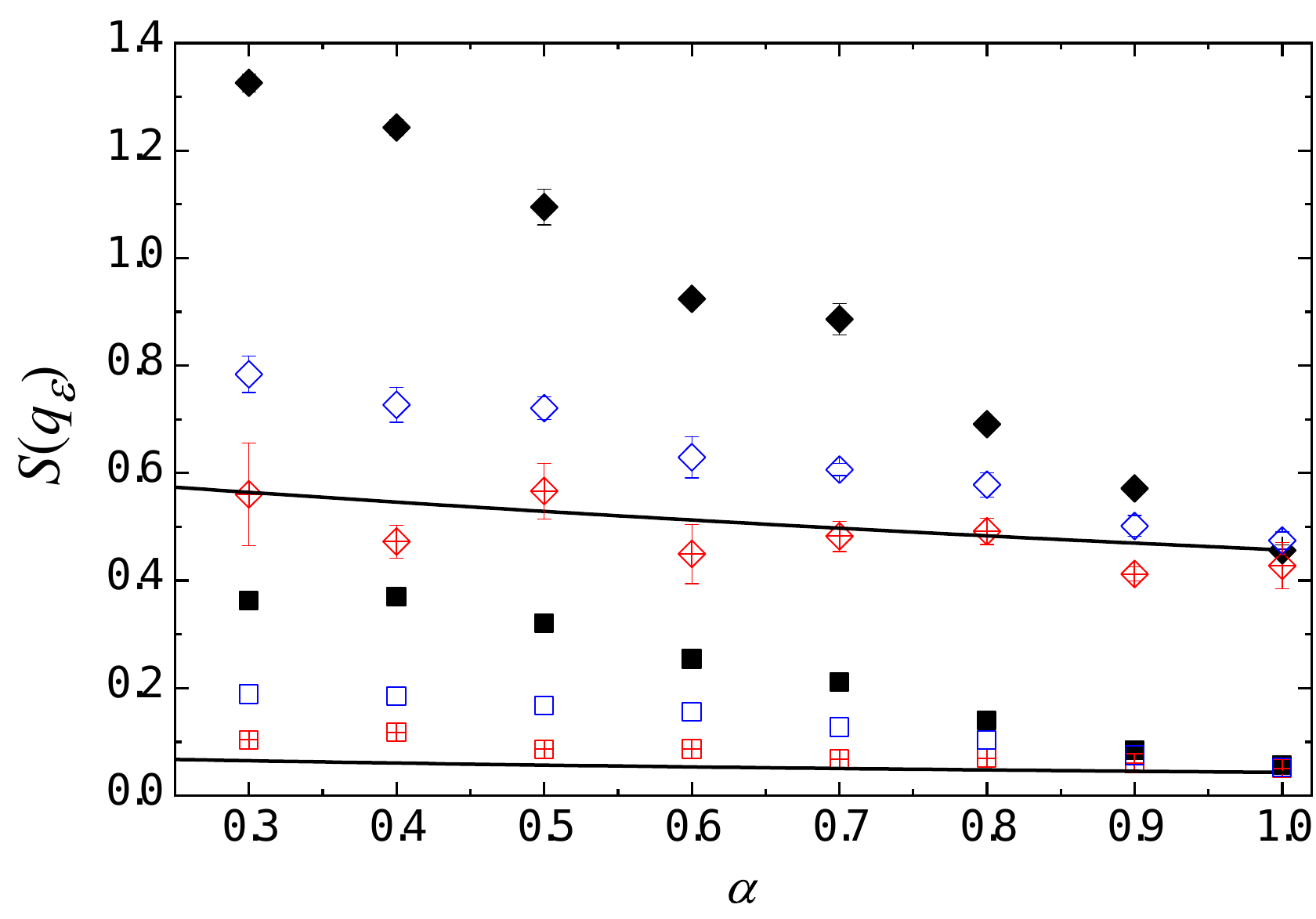}
	\end{center}
   \caption{Simulation values of $S(q_\varepsilon)$
   versus $\alpha$ for $\phi=0.1$ (diamonds) and $0.4$ (squares), with $\gamma^*=0.2$ (filled symbols), 1 (open symbols), and 5 (crossed open symbols).  
   For $\phi=0.1$,  $q_\varepsilon\sigma=0.256$, whereas $q_\varepsilon\sigma=0.287$ for $\phi=0.4$. 
   {The lines show  the theoretical prediction for  $S(0)$, given by Eq.~\eqref{kappaS0}, with $\hat{\kappa}$ given by Eq.~\eqref{hatkappa},} 
   for $\phi=0.1$ (top line) and $0.4$ (bottom line).
  { Error bars are not shown for $\phi=0.4$ because they are smaller than the symbols.}
   }
	\label{fig:S0Sigma}	
\end{figure}

At higher density, $\phi=0.4$, see Fig.~\ref{fig:Sqgamma02-5phi04}, the same qualitative trends persist, although the dependence on $\gamma$ becomes noticeably weaker, in accordance with the results for $g(r)$. For inelastic cases, the simulation results still show some sensitivity to the drag coefficient, but this effect is now only slight, even though it remains more visible in the small-$q$ region and becomes stronger as inelasticity increases. In particular, $S(0)$ depends only weakly on $\gamma$, especially for weak inelasticity.
(see Fig.~\ref{fig:S0Sigma}).
In this denser regime, the main deficiency of the PY approximation is no longer concentrated only at small $q$, but also becomes apparent in the region of the first maximum, where it departs more clearly from both the simulation data and the RFA as $\alpha$ decreases. The RFA also deteriorates with increasing inelasticity, but it still provides a reasonable overall description, with the largest discrepancies with respect to the simulation results appearing, at most, for low friction in the small-$q$ region and near the first peak.

Outside the smallest-$q$ region, the RFA remains the most accurate of the two equilibrium-like descriptions and reproduces the simulation data rather well over the rest of the explored range. The upward deviation of the simulation results at small $q$, namely, the enhancement of $S(q)$ with respect to both RFA and PY, should therefore be interpreted not as a failure of the particle-scale structural theory, but as a distinct nonequilibrium contribution associated with long-wavelength correlations of hydrodynamic origin, which are characteristic of driven granular media \cite{Cecconi2004,Puglisi2012,Noije1999}. Since such correlations are absent in equilibrium-fluid theories with elastic molecular collisions, they cannot be captured by theoretical expressions for $S(q)$ derived within that framework.
{Recall that our theoretical estimate of $S(0)$ through $\hat{\kappa}$ relies on Eq.~(10), which is strictly an equilibrium relationship.}

For large $\gamma^*$, the low-$q$ enhancement becomes less pronounced, a trend that is consistent with the idea that drag suppresses long-wavelength nonequilibrium correlations {\cite{Puglisi2012}}. As a result, 
the simulation results lie closer to the PY and RFA predictions. 
By contrast, fluctuating-hydrodynamic theory for randomly
driven granular fluids without drag predicts 
$S(q)\sim q^{-2}$ at small wave number \cite{Noije1999}. 
The smallest wave numbers accessible in our simulations are $q_{\min}\sigma\approx 0.2$ (see below). Consequently, the simulations cannot fully establish the asymptotic behavior as $q \to 0$ and  exclude a singularity. Nevertheless, the available low-$q$ data are consistent with a finite 	small-$q$ limit. Specifically, taking the RFA prediction 	as an approximate structural background, the excess 	$\Delta S(q)=S_{\mathrm{simulation}}(q)-S_{\mathrm{RFA}}(q)$ is compatible with the Lorentzian  $\Delta S(q)\sim 1/(q^2+q_c^2)$, whereas a $q^{-2}$ dependence is incompatible with the data over the same low-$q$ interval. (The above Lorentzian form agrees with the one reported by Cecconi et al. \cite{Cecconi2004} for the low-$q$ behaviour of a 	one-dimensional granular fluid coupled to a Langevin bath). In summary, our data support, but do not 	establish, finiteness of the enhancement for $q\to0$. 
Anyway, a detailed  analysis of this enhancement lies outside the objective of the present work, which is to provide a simple analytical description of the short- and intermediate-range structure of the suspension.

To compare theory and simulation of $S(q)$ for small $q$  more directly, Fig.~\ref{fig:S0Sigma} shows the values of $S(q_\varepsilon)$, 
where $q_\varepsilon$ is the smallest wave number  retained in the analysis, as a function of $\alpha$  for $\phi=0.1$ and $\phi=0.4$ and for several values of $\gamma^*$. Smaller wave numbers were excluded in the analysis because of their large statistical uncertainties. 
Thus, $S(q_\varepsilon)$ serves as a finite-wavenumber proxy for $S(0)$. 
The corresponding theoretical equilibrium-like values of the small-$q$ limit, $S(0)=\hat{\kappa}$, obtained from Eqs.~\eqref{hatkappa} and \eqref{kappaS0}, are also shown for both densities. 
In contrast to the behavior of $g(\sigma)$ (see Fig.~\ref{fig:gSigmaBig}), the deviations between theory and simulation in this long-wavelength sector are quite large at small $\gamma^*$. This indicates that the theoretical approximations affect $g(r)$ and $S(q)$ differently, i.e., they have a different impact on short-range structure and long-wavelength correlations.

In summary, at the level of $S(q)$, the equilibrium-like description proposed in this article becomes clearly more accurate as the density of the suspended grains increases, whereas it deteriorates as the inelasticity increases. In the elastic case, both PY and RFA are in good agreement with the simulation data, which, within simulation accuracy, are independent of the friction coefficient. For inelastic systems, the dependence on friction becomes appreciable mainly at low density and high inelasticity, and it shows up most clearly in the small-$q$ region. In that regime, the simulations exhibit a strong enhancement of $S(q)$ at low $q$ relative to the equilibrium-like RFA and PY predictions, especially for weaker friction. This low-$q$ enhancement decreases significantly as the density increases. Overall, RFA provides a clearly better description than PY as the inelasticity increases, although both approaches show their main limitations for small-$q$ values and, at high density, also in the region of the first peak.  

\section{Conclusions}
\label{sec:Conclu}
 
The main result of this work is that an equilibrium-inspired RFA, supplied with nonequilibrium input for $g(\sigma)$ and for the reduced isothermal susceptibility $\hat{\kappa}$, provides a very good description of the short- and intermediate-range structure of granular suspensions when compared with simulations in which the dynamics are  modeled by a generalized Langevin equation with drag, stochastic forcing, and inelastic hard-sphere collisions. 
In this effective description, the action of the surrounding fluid is represented through drag and stochastic forcing, without explicit hydrodynamic interactions.
We find that the RFA reproduces the simulation results for $g(r)$ much more accurately than the PY approximation  when collisions are inelastic, and it also yields a good account of $S(q)$ outside the smallest-$q$ region, where additional nonequilibrium long-wavelength fluctuations become important. This shows that an equilibrium-based hard-sphere approach can remain useful for describing the structure of a nonequilibrium suspension without attempting to account explicitly for all nonequilibrium contributions.

This viewpoint is consistent with the more general idea that equilibrium-like descriptions of structure can still be useful in nonequilibrium hard-core systems, as already suggested by Lutsko's analysis of the homogeneous cooling state of a granular gas { and by Nichol and Daniels  \cite{Nichol2012}   
in their analysis of the equipartition of energy between the translational and rotational degrees of freedom in a driven dense granular gas.}    From a practical point of view, the appeal of the RFA lies not only in its accuracy but also in its simplicity. Once the required input parameters are known, the theory provides fully explicit analytical expressions for $S(q)$ and for $g(r)$ in Laplace space. This makes it a simple and physically transparent framework for describing the structural properties of the suspension.

Our simulation results show that the particle-scale structure is controlled by the combined effects of inelasticity, drag, and density. In the elastic limit, the structure is essentially insensitive to drag, whereas increasing collisional inelasticity makes both $g(r)$ and $S(q)$ more sensitive to drag. This dependence is much stronger at low density than at moderate density. Over a broad range of conditions, well beyond the elastic regime, the RFA provides a useful and globally accurate description of the simulation data. Moreover, for inelastic collisions, the improvement of the RFA prediction for $g(r)$ over the PY approximation is significant. Indeed, the PY approximation gives a clearly poor description of $g(r)$ near contact in inelastic cases, and its performance deteriorates further as the collisions become more inelastic.

At the same time, the simulations show a clear increase of $S(q)$ over the equilibrium-like RFA prediction in the small-$q$ region, especially at low density, strong inelasticity, and weak drag. This enhancement is associated with the appearance of long-wavelength nonequilibrium correlations, which are a characteristic feature of driven granular systems. For this reason, the present theory should be viewed mainly as a description of the short- and intermediate-range structure of a homogeneous, isotropic steady state, while the small-$q$ enhancement points to additional collective physics at larger length scales. In other words, the low-$q$ behavior is not simply a deficiency of the structural theory, but rather the signature of a different large-scale contribution superimposed on a shorter-range structure that is otherwise well described by the RFA.

Our results suggest that the RFA may also be useful in other nonequilibrium states whose short- and intermediate-range structure remains close to that of equilibrium hard-sphere fluids. {For example, one could consider a granular suspension of rough inelastic hard spheres, i.e., a system where the rotation of the grains is taken into account \cite{Jenkins2002, Kumaran2006a, Ruben2020}}. 
Finally, a natural next step is to extend the analysis to multicomponent granular suspensions. Such an extension appears feasible because an explicit RFA formulation already exists for additive hard-sphere mixtures, with contact values and isothermal compressibility as basic inputs, while kinetic-theory descriptions for multicomponent granular suspensions can provide approximate nonequilibrium expressions for the quantities required by the theory.

\acknowledgments
S.B.Y.  acknowledges financial support from Grant No. PID2024-156352NB-I00 funded by MCIU/AEI/10.13039/501100011033/FEDER, UE and from Grant GR24022 funded by Junta de Extremadura (Spain) and by ERDF ``A way of making Europe''. 
A.M.P. also acknowledges financial support from Grants No. PID2021-127836NBI00, funded by MCIN/AEI/10.13039/501100011033/FEDER ``A way to make Europe'', and PID2025-170983NB-I00.

\section*{Data Availability}
The data supporting the findings of this study are openly available \cite{datosEnZenodo}.




%

\end{document}